\documentclass{article}

\usepackage[english]{babel}
\usepackage{ulem}
\usepackage[letterpaper,top=2cm,bottom=2cm,left=3cm,right=3cm,marginparwidth=1.75cm]{geometry}

\usepackage{amsmath}
\usepackage{graphicx}
\usepackage[colorlinks=true, allcolors=blue]{hyperref}
\usepackage[dvipsnames,table,xcdraw]{xcolor} % just for color of comments

\title{How to survive the Squid Games using probability theory.}
\author{Elena Moltchanova \thanks{School of Mathematics and Statistics, University of Canterbury, Private Bag 4800, Christchurch 8140, New Zealand}, Miguel Moyers-Gonz\'alez.\footnotemark[1], Geertrui Van de Voorde\footnotemark[1],\\ Jos\'e Felipe Voloch\footnotemark[1], Philipp Wacker.\footnotemark[1] }
\date{} 
\begin{document}
\maketitle

\begin{abstract}
In this paper, we consider how probability theory can be used to determine the survival strategy in two of the ``Squid Game" and ``Squid Game: The Challenge" challenges: the Hopscotch and the Warships. We show how Hopscotch can be easily tackled with the knowledge of the binomial distribution, taught in introductory statistics courses, while Warships is a much more complex problem, which can be tackled at different levels.
\end{abstract}

\section{Introduction.}

The Korean Netflix series ``Squid Games" and its reality show spin-off ``Squid Games: The Challenge" have attracted an unexpectedly large audience worldwide, and have resulted in multiple academic papers. (The search for ``squid games" on Google Scholar produces about 270 results). However, most papers are concerned with socioeconomic and psychological aspects of the show, and we found only one paper referring to more mathematical and game-theoretic aspects of the show \cite{geerling2023using}.

There are at least two challenges (the ``Hopscotch" in the original show, also referred to as ``Glass Bridge" in the reality spin-off,  and the ``Warships" in the reality spin-off), which lend themselves explicitly to analysis via the probability theory. And while some people might say that they would rather die than do statistics, we have decided to explore how statistics can be used to help survive the Squid Games.

While our treatment of the problems is firmly tongue-in-cheek, we discovered quite a few delightful tangents in the process of discussing them. While the ``Hopscotch"/``Glass Bridge" is a very simple problem, and can certainly be used as a nice example in introductory probability theory courses, the Warships is very different and can easily be a topic for more serious study. 

\section{Panem et circenses.}

\subsection{Hopscotch.}

In the glass bridge challenge, the participants have to cross a bridge, consisting of two rows of seventeen glass tiles. The glass tiles are side-by-side, and out of each pair, one is made of tempered glass, and another of fragile glass, which shatters when stepped upon and sends the player either to their death (in the original ``Squid Game" show) or to their elimination (in the spin off ``Squid Game: The Challenge"). The players pick their order number before the game begins, and go one-by-one with the objective of crossing the bridge safely. Once the first person is eliminated, the second follows in their footsteps, and so on.

In the above set-up, the probability that the first person completes seventeen steps without ever stepping on a fragile tile is 

\begin{equation*}
\left(\frac{1}{2}\right)^{17} \approx 7.6\times 10^{-6},
\end{equation*}

i.e., vanishingly small. The players with the order numbers $18$ and above, on the other hand, are obviously safe, because even if the first 17 players always choose the wrong tile, the path will be completely ``discovered" by them. What are the probabilities of survival for the players in between? And which numbered player has the highest chance of crossing the bridge first?

Let $n$ be the total number of steps, and the probability of elimination in each step be $\frac{1}{m}$. The number of missteps $x$ then has a binomial distribution: 

\begin{equation*}
x \sim \text{Binom}(n,1/m).
\end{equation*}

The probability that player $k$ is the first to make it across the bridge is equivalent to exactly $k-1$ players having made a misstep on the bridge:

\begin{equation}
\label{eq:bin.m1}
Pr(\text{player }k\text{ finishes first})=Pr(x=k-1) = \binom{n}{k-1}\left(\frac{1}{m}\right)^{k-1} \left(1-\frac{1}{m}\right)^{n-(k-1)},
\end{equation}

and the probability of survival for the player $k$ is then

\begin{equation}
\label{eq:bin.m2}
Pr(k \text{ survives}) = Pr(x < k) =\sum_{x=0}^{k-1} \binom{n}{x}\left(\frac{1}{m}\right)^{x} \left(1-\frac{1}{m}\right)^{n-x}.
\end{equation}

The expected number of the first player crossing the bridge is then 

\begin{equation*}
1+E(x)=1+\frac{n}{m}.
\end{equation*}

\begin{figure}
    \centering
    \includegraphics[width=0.8\linewidth]{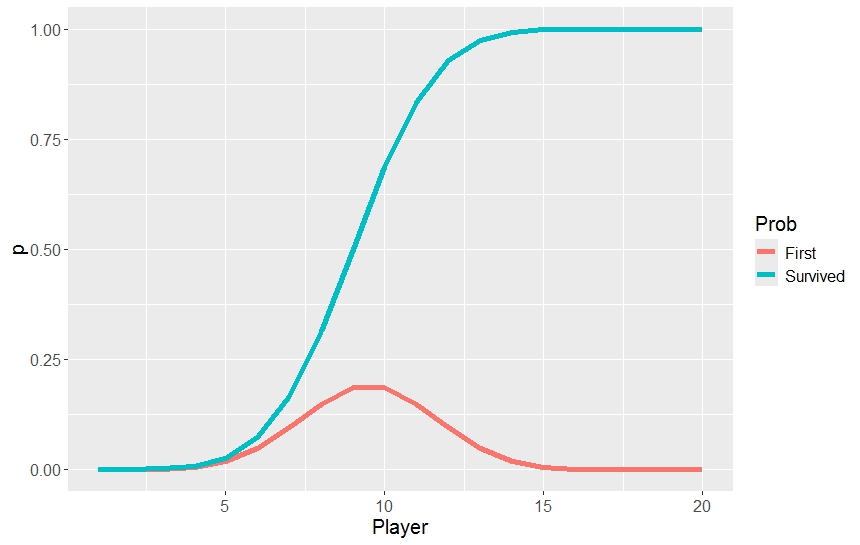}
    \caption{Probabilities of survival and being the first to cross the bridge in the Hopscotch challenge, when the number of steps $n=17$ and the probability of a misstep $m=1/2$.}
    \label{fig:hopscotch}
\end{figure}

For $n=17$ and $m=2$ in the show, this equals $9.5$. (This also happens to be the average of the two modes of the Binomial distribution $\text{Binom}(17, 0.5)$ at $k=9$ and $k=10$. The probabilities of survival and of being the first to cross the bridge for this case are shown in Figure \ref{fig:hopscotch}. It shows that if you want increase your chances of survival, you want to be in the tail of the queue, while if you want to maximize your chances of being the first to cross the bridge, you want to be number 9 or 10.

\subsection{Warships.}

Another game in Squid Game: The Challenge is ``Warships", which is a modified version of the game ``Battleships". In this game, two teams face off against each other. Each team has a captain, a lieutenant, and $13$ sailors, placed into $4$ ships of lengths of $5$, $3$, $3$, and $2$ respectively. As in the Battleships game, each team decides where to place their ships, and the teams can only see their part of the sea, which is $8\times 8$ squares.

The captain and lieutenant are responsible for the strategy: they decide where to aim the missiles when shooting at the other team. If they hit a ship, they get another go. Otherwise, the turn passes on to the other team.

The first team to completely sink two of the enemy's ships wins, and the captain and the lieutenant of the losing team are eliminated. Also, any sailors on a completely sunk ship are eliminated.

Who goes first is decided randomly after the roles have been assigned.

In the show, there was quite a bit of argument about whether it is better to be the command, i.e., either captain or lieutenant. Another interesting question is whether being on a larger ship gives you a higher probability of survival than being on a smaller ship, since, intuitively, smaller ships are harder to find but faster to sink. This means it is interesting to investigate the following questions:
\begin{enumerate}
    \item If you want to maximise your chances of survival, should you choose to be \textit{command} or \textit{sailor}?
    \item If you are a sailor and want to maximise chances of survival, what is the length of the ship (5, 3, or 2) that you should board?
\end{enumerate}

Let's take a look at these two questions in turn. Note that, while in the show various players claimed outstanding knowledge of strategy as well as psychological insights into the enemy's strategic thinking, we are mostly going to assume that every player is as good as any other in their ability to figure out the optimal strategy for the game. We also assume that the ships are placed on the board randomly according to the rules.

Let's begin by determining the probability of survival for the captain and the lieutenant. Since no team has an advantage to begin with, the probability of winning the game is $\frac{1}{2}$, and so is the probability of the command surviving. Now, let's turn to the probability of surviving as a sailor.

In general, we can state that

\begin{eqnarray}
\label{eq:BaseFormula}
Pr(\text{being on a sunk ship}) &=& Pr(\text{losing team})Pr(\text{being on first or second sunk boat in losing team's fleet})+\nonumber\\
& &Pr(\text{winning team})Pr(\text{1 ship sunk}|\text{winning team})\times\nonumber\\
& &Pr(\text{being on the only sunk ship of winning team}|\text{winning team, 1 sunk ship})\nonumber\\
&=&\frac{1}{2}p_{12}+\frac{1}{2}(1-\pi)p_1,
\end{eqnarray}

where $p_1$ is the probability of being on the first ship sunk, $p_{12}$ is the probability of being on either of the first two ships sunk, and $\pi$ is the probability of the winning team not losing a single ship.

To begin with, consider the simplified situation where all four ships are of the same length, and are thus equally likely to be sunk in any order so that $p_1 = 1/4$ and $p_{12}=1/2$. Equation \ref{eq:BaseFormula} then becomes:

\begin{eqnarray*}
Pr(\text{being on a sunk ship}) &=& \frac{1}{2}\frac{1}{2}+\frac{1}{2}(1-\pi)\frac{1}{4}\\
&=&\frac{1}{8}(3-\pi) < \frac{3}{8} < \frac{1}{2}.
\end{eqnarray*}

Thus, when all four ships are identical, the probability of surviving as a crew member is quite a bit better than the probability of surviving as a captain or a lieutenant. When the ships are not identical, the smaller than average ship will have a lower probability of being among the first two sunk, and thus the probability of its crew members' elimination will be even lower. On the other hand, if a ship is virtually guaranteed to be the first one to be sunk, the probability of elimination becomes

\begin{eqnarray*}
Pr(\text{being on a sunk ship}) &=& \frac{1}{2}+\frac{1}{2}(1-\pi)\\
&=&\frac{1}{2}(2-\pi) \geq \frac{1}{2},
\end{eqnarray*}

i.e., higher than that for the command.

So, to answer our question, we need to figure out the parameters $p_1$, $p_{12}$, and $\pi$. While producing an algorithm for playing ``Warships" more or less effectively is far beyond the scope of this study, we intend to use a combination of basic probability theory and some simplified simulations to estimate the parameters we need to solve this problem.

\subsubsection{Placing the Ships.}

There are $8\times(9-L)\times2$ ways to place a ship of length $L$ on the board. That gives $64$, $96$, and $112$ options for $L=5$, $3$, and $2$ respectively. Because there may be possible overlaps, and because the two $L=3$ ships are exchangeable, the number of legitimate arrangements is substantially smaller than $64\times96\times96\times112=66, 060, 288$. We have produced a list of all possible arrangements for each ship, and wrote an algorithm to retain only the non-overlapping arrangements. Not making a distinction between the two ships of size $3$, this results in $28, 876, 784$ distinct placements. 

Arguably, placing the ships in adjacent cells makes it easier for the opponent to inadvertently find the second ship while trying to sink the first one. A more clever strategy may thus be to place ships so that no two ships are adjacent by side. The number of such ``buffered" arrangements is smaller at $6, 406, 464$.

\begin{figure}
\centering
\includegraphics[width=1\linewidth]{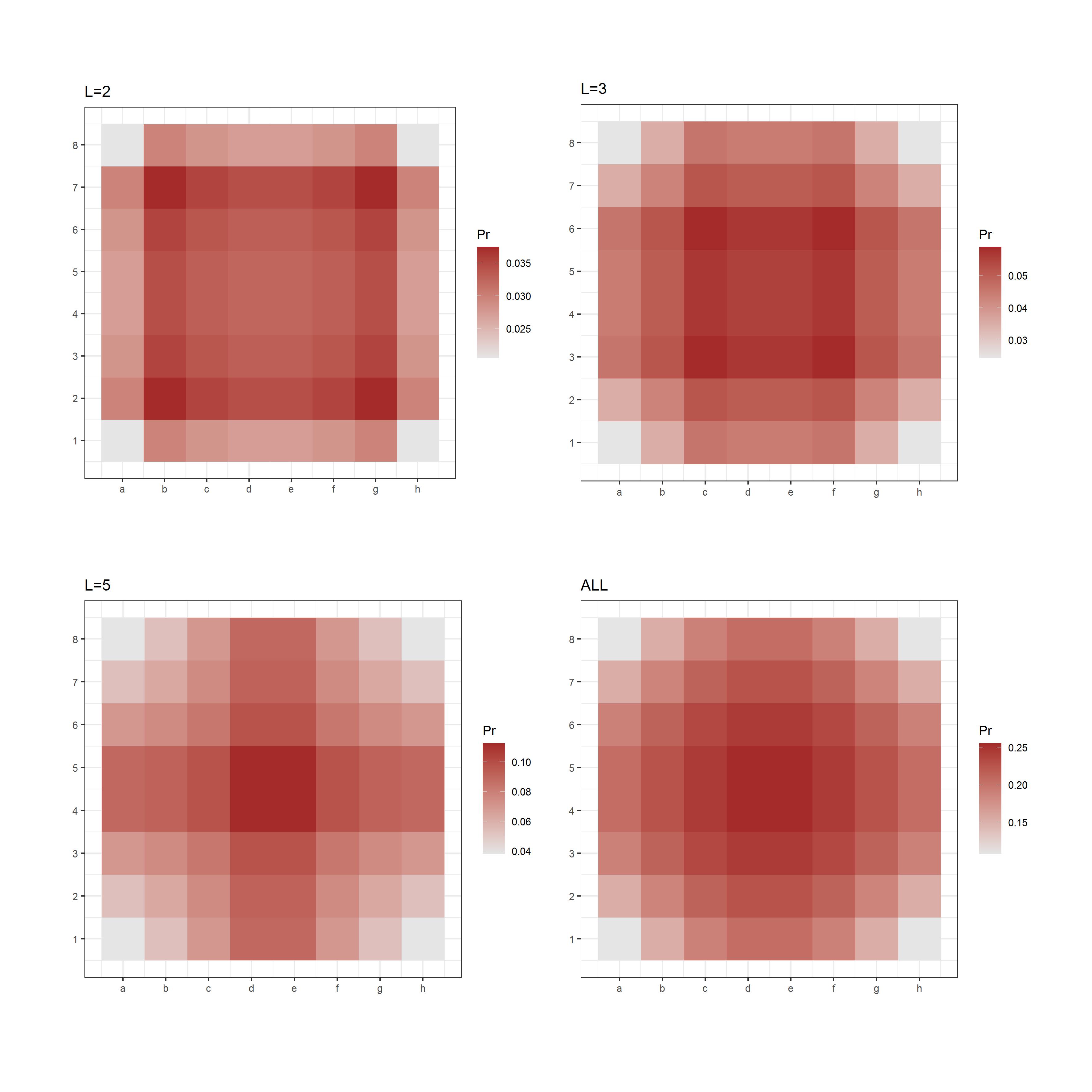}
\caption{\label{fig:placement}Probability of a ship being located in each cell under the standard rules. The central cells are more likely to contain a ship.}
\end{figure}

\begin{figure}
\centering
\includegraphics[width=1\linewidth]{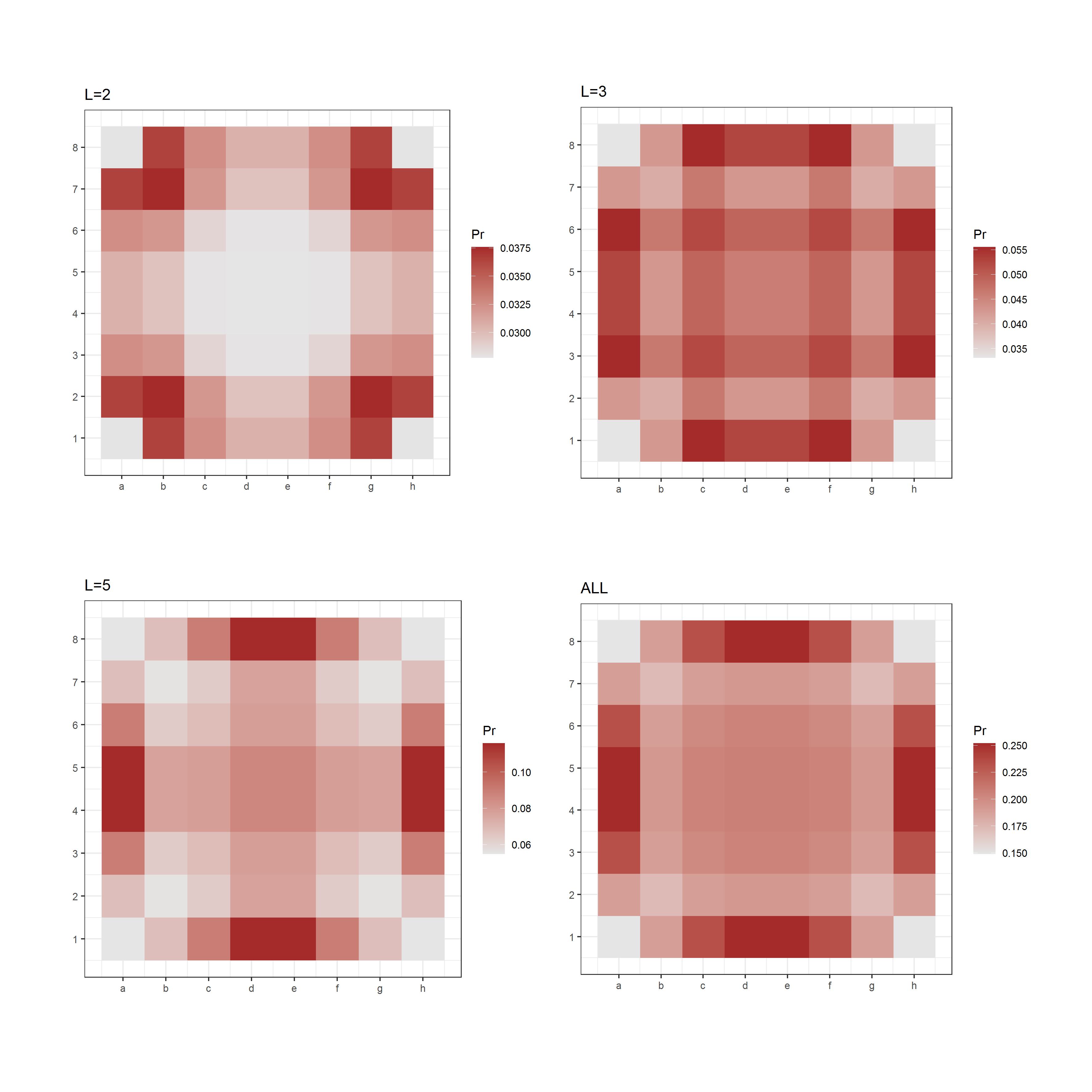}
\caption{\label{fig:placementbuff}Probability of a ship of length $L$ being located in each cell when only allowing buffered arrangements. Note that the central cells are less likely to contain a ship.}
\end{figure}

It is interesting to take a look at the probability that a cell will have a ship located on it for the two arrangements described above. These probabilities are shown in Figures \ref{fig:placement} and \ref{fig:placementbuff}.

\subsubsection{Hunting for the ship.}\label{sss:hunting}

\begin{figure}
\centering
\includegraphics[width=0.5\linewidth]{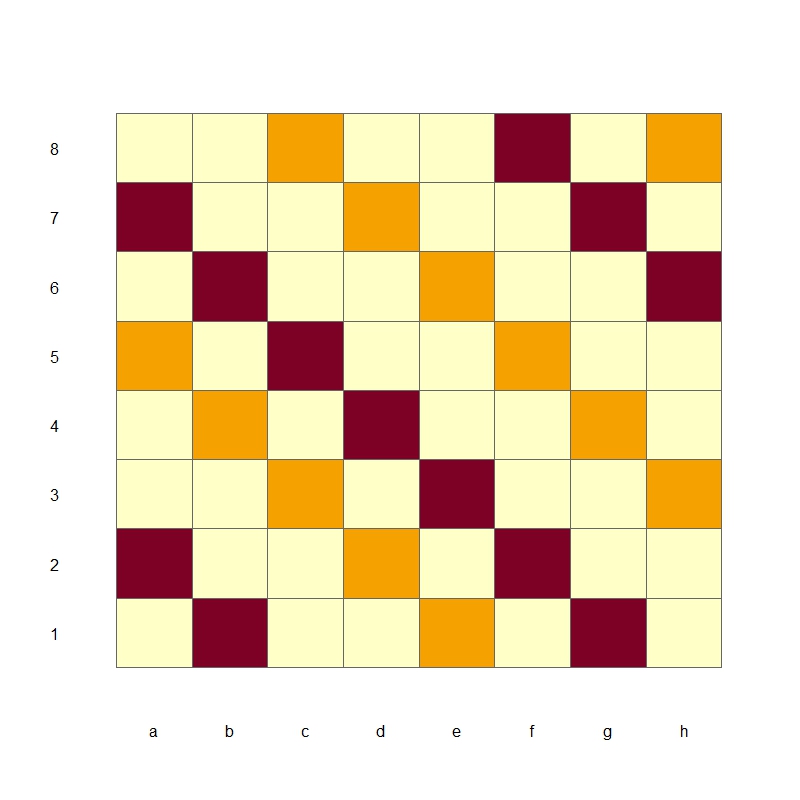}
\caption{\label{fig:grid5}A grid, guaranteed to find a ship of length $5$ within 12 moves (orange cells), and another ship in further 12 moves (yellow cells).}
\end{figure}

The problem of the optimal strategy in `Battleships' has been addressed before, by, for example, Rodin et al, \cite{rodin1988developing}, and more recently by  \cite{crombez2020efficient},\cite{audinot2014optimal}. Predictably, there have also been some forays into reinforcement learning (see \cite{clementis2013supervised} and \cite{kancko2020reinforcement}). While discussing the optimal strategy for either placing the ships or hunting is beyond the scope of this study, we are going to consider four possible strategies for finding a ship:

\begin{itemize}
    \item {The ``Diagonal" strategy} involves hitting cells highlighted in Figure \ref{fig:grid5}, first all the orange cells, then all the brown cells. This strategy guarantees finding the $L=5$ ship within $12$ turns, and the second ship within additional $12$ turns.
    \item{The ``Smart Diagonal" coverage-proportionate strategy}, where the cells highlighted in Figure \ref{fig:grid5} are hit in order of decreasing probability of a ship being located there as per Figures \ref{fig:placement} and \ref{fig:placementbuff}.
    \item{The ``Regular" strategy},
    where the cells in the entire field are hit in order of decreasing probability of a ship being located there as per Figures \ref{fig:placement} and \ref{fig:placementbuff}.
    \item{The ``Completely Random" strategy}, where the cells are hit completely at random.
\end{itemize}

\subsubsection{After the first hit.}

% once the ship is hit.
How many additional turns does it take to sink a ship, once it is hit? Keep in mind, that if the team has successfully hit a ship, it is allowed to go again. So the number of turns equates the number of misses. For a ship of length $L=2$, assuming that it is not at the edge of in the corner of the sea and assuming that no other ships are adjacent to it,  the remaining square can be found immediately with probability $1/4$, or after one miss with probability $3/4\times1/3=1/4$, or after two misses with probability $3/4\times2/3\times1/2=1/4$, or after three misses with the remaining probability $1/4$. For bigger ships, enumerating all the paths to destruction yields the probability distributions presented in Table \ref{tab:ProbSink}. 

\begin{table}
    \centering
    \begin{tabular}{l|c|c|c}
         misses, $x$ & L=2 & L=3 & L=5 \\
         \hline
         0 & 1/4 & 3/18& 1/10\\
         1 & 1/4 & 6/18 & 4/10\\
         2 & 1/4 & 5/18 & 3/10\\
         3 & 1/4 & 4/18 & 2/10\\
         \hline
         $E(x)$&  1.50 & 1.56 & 1.60 \\
    \end{tabular}
    \caption{Probability distribution for the number of turns to sink a ship of size $L$, located in the middle of the sea, once it is hit.} 
    \label{tab:ProbSink}
\end{table}

We see that the average time taken to sink the ship does not substantially vary with the size of the ship. While the logic of finishing the ship appears intuitive to a human played, implementing an efficient algorithm is far from trivial, and having possibly adjacent ships only complicates it further. 

\subsubsection{Simulation Studies.}

After listing all possible ship deployments under the ``no-buffer" and ``buffer" settings, we have used simulations studies to see how long it will take to hit the first ship using each of the four strategies explained in section \ref{sss:hunting}. Avoiding the issue of finishing the ship off, we then assumed that the hunt would continue until the second (different) ship was hit. The probabilities of each of the ships being hit first or within the first two are shown in Table \ref{tab:ProbSink}. Note, that they do not sum up to one, because we have two ships of length $L=3$.

One way to determine the probability of not a single ship being sunk on the winning side, $\pi$ is to simulate a lot of games, with two games playing each other, and evaluate the frequency of the event in question. However, assuming that the strategy of one team is not in any way influenced by the strategy of the other team, one can use a one-sided simulation to obtain the distribution for the number of shots it takes to hit the first ship ($t_1$), and the (total) number of shots it takes to hit the first two ships ($t_2$). Assuming it takes $\delta_1$ turns to sink the first ship that was hit, and $\delta_2$ turns to sink the second ship that was hit, the probability of not losing any ships on the winning side is then  $\pi=Pr(t_1+\delta_1 > t_2+\delta_1+\delta_2)=Pr(t_1 > t_2+\delta_2)$, and can also be obtained from simulations.

\begin{figure}
\centering
\includegraphics[width=0.8\linewidth]{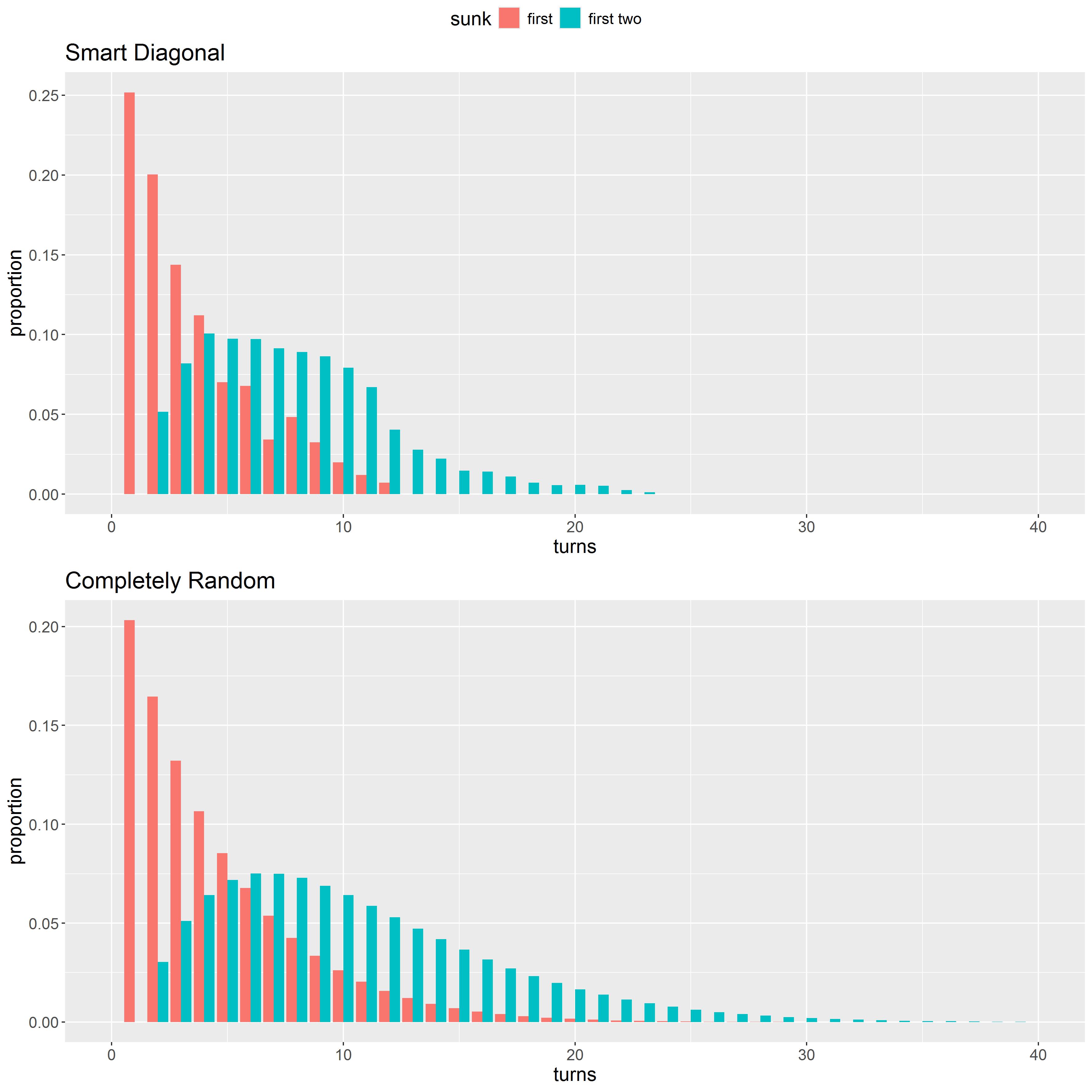}
\caption{\label{fig:times}Observed distribution for the number of turns until the first hit, and the second hit for the ``Smart Diagonal" and the``Completely Random" strategies respectively.}
\end{figure}

The summary statistics for time $t_1$ and $t_2$ under different strategies are shown in Table \ref{tab:times}. As postulated, the ``Smart Diagonal" strategy produces the fastest results. Figure \ref{fig:times}

Using the information in Table \ref{tab:ProbSink}, we have used the average estimate $\delta_t=1.60$ to obtain $\pi_{\text{ave}}$. We have also used an extremely conservative $\delta_t=0$ (assuming that ships are sunk without any misses) to provide an upper bound $\pi_{\max}$ for the probability of no sunk ship by one team while the other sinks both. The results are shown in Table \ref{tab:probs_intermediate}.

Note, that for completely random shooting, the probability that the first ship is of length $L$ is $L/13$, where $13=5+3+3+2$ is the sum of the lengths of all the ships. This yields $0.154$, $0.231$, and $0.385$ for $L=2, 3$, and $5$ respectively, and agrees with the estimates for random hunting whether with or without placement buffer in Table \ref{tab:probs_intermediate}. Also notice that because the ``Diagonal" strategies are designed to find the longest ship first, the simulated probabilities of $L=5$ being hit first are higher than that for the ``Completely Random" and ``Regular" strategies.

Substituting the estimates from Table \ref{tab:probs_intermediate} into Equation \ref{eq:BaseFormula} yields the probabilities of a ship of length $L$ being eliminated, which are reported in Table \ref{tab:probs_final}. The probability of the ships of lengths $L=2$ and $L=3$ being eliminated are firmly below $0.5$. Thus, it is better to be on those smaller ships than to be a captain or a lieutenant. For the longest ship, $L=5$, the probability of being eliminated is higher than $0.5$, except for the Regular strategy.

\begin{table}
    \centering
    \begin{tabular}{l|ccc|ccc|c|c}
        &$p_1$ &&&$p_{12}$ & &&&\\
            Strategy &L=2   &L=3   &L=5   &L=2   &L=3   &L=5    &$\pi_{ave}$&$\pi_{\max}$\\
               \hline
Buffer&&&&&&&&\\
\hline
Completely Random         &0.154 &0.231 &0.385 &0.342 &0.486 &0.685  &0.123     &0.159\\
Regular        &0.146 &0.237 &0.379 &0.346 &0.516 &0.622  &0.080     &0.118\\
Diagonal  &0.121 &0.207 &0.465 &0.220 &0.385 &0.662  &0.045     &0.103\\
Smart Diagonal &0.108 &0.201 &0.490 &0.220 &0.385 &0.662  &0.147     &0.150\\
\hline
No Buffer&&&&&&&&\\
\hline
Completely Random        &0.154 &0.231 &0.385 &0.342 &0.486 &0.685  &0.123      &0.159\\
Regular        &0.167 &0.249 &0.334 &0.390 &0.519 &0.571  &0.125      &0.146\\
Diagonal  &0.125 &0.210 &0.455 &0.224 &0.378 &0.665  &0.102      &0.146\\
Smart Diagonal &0.122 &0.213 &0.452 &0.224 &0.378 &0.665  &0.0095      &0.140\\
\hline
    \end{tabular}
    \caption{Probability of being sunk first or being among the first two ships sunk under different placement and hunting scenarios, as well as the probabilities of no ships being sunk by one team while the other sinks both under the average $\delta_t=3.06$ and extremely conservative $\delta_t=0$ assumptions.}
    \label{tab:probs_intermediate}
\end{table}

Putting these numbers into Equation \ref{eq:BaseFormula} results in the probabilities of elimination reported in Table \ref{tab:probs_final}.

\begin{table}
    \centering
    \begin{tabular}{c|ccc|ccc}
    Strategy & &given $\pi_{ave}$ &&&given $\pi_{max}$ &\\
    \hline
               &L=2     &L=3     &L=5     &L=2     &L=3     &L=5\\
               \hline
Buffer&&&&&&\\       
\hline
Completely Random         &0.239 &0.344 &0.511 &0.236 &0.340 &0.504\\
Regular        &0.238 &0.362 &0.476 &0.236 &0.359 &0.473\\
Diagonal  &0.164 &0.286 &0.540 &0.161 &0.281 &0.529\\
Smart Diagonal &0.158 &0.282 &0.550 &0.156 &0.278 &0.539\\
\hline
No Buffer&&&&&&\\
\hline
Completely Random         &0.239 &0.344 &0.511 &0.236 &0.340 &0.505\\
Regular        &0.268 &0.369 &0.432 &0.266 &0.366 &0.428\\
Diagonal  &0.169 &0.283 &0.537 &0.166 &0.279 &0.527\\
Smart Diagonal &0.167 &0.285 &0.537 &0.164 &0.281 &0.527\\
\hline

    \end{tabular}
    \caption{Probabilities of being eliminated on a ship of length $L$ under different placement and elimination strategies as well as different assumptions about the time taken to finish a ship.}
    \label{tab:probs_final}
\end{table}

\begin{table}
    \centering
    \begin{tabular}{l|ccc|ccc}
        &&$t_1$&&&$t_2$&\\
            Strategy & mean   &median   &max   &mean   &median   &max \\
               \hline
Buffer&&&&&&\\
\hline
Completely Random   &4.6 & 4    &44 &10.5 &9    &53\\
Regular             &4.9 & 3    &39 &12.4 &11   &47\\
Diagonal            &4.0 & 3    &12   &8.3 &8    &23\\
Smart Diagonal      &3.6 & 3    &12   &7.8 &7    &23\\
\hline
No Buffer&&&&&&\\
\hline
Completely Random   &4.6  &3&42&10.5& 9&54\\
Regular             &5.7  &3&39&14.9& 13&47\\
Diagonal            &3.9  &3&12&8.1&  8&23\\
Smart Diagonal      &3.5  &3&12&7.6&  7&23\\
\hline
    \end{tabular}
    \caption{The number of turns until the first hit, $t_1$, and until the second hit, assuming the first ship is sunk immediately, $t_2$ under different placement and hunting scenarios.}
    \label{tab:times}
\end{table}

\section{Conclusions}

In this paper, we have considered the Hopscotch and the Warships challenge in the Squid Games and its spin-off show. We have explained how the probability of survival and of being the first over the bridge for the Hopscotch challenge can be easily evaluated using the binomial distribution. Since the binomial distribution is often taught in first year introductory classes, this may be a nice example to engage students.

The Warship challenge is much more complex. Although, we have stopped short of producing an optimal algorithm based on reinforcement learning, we have managed to get quite far within a simpler set-up. For example, the solution for a special case of four identical ships is very obvious once known but may require some thinking to arrive at. Also, while the efficient strategy of finishing off a ship may seem fairly obvious to a human, explaining it to a computer turns out to be a much harder task. This makes reinforcement learning or similar a very tempting option, which may be an appropriate topic for a postgraduate course project.

We have explored two different placement strategies and four different hunting strategies, but our list is obviously not exhaustive. Other strategies may be compared. Furthermore, while in the Squid Game: The Challenge, the game only happens once, in the repetitive environment, you can imagine the two players mutually and continuously adjusting their own placement and hunting strategies depending on what the other is doing.

\bibliographystyle{alpha}
\bibliography{sample}

\newcommand{\etalchar}[1]{$^{#1}$}
\begin{thebibliography}{GNR{\etalchar{+}}23}

\bibitem[ABV14]{audinot2014optimal}
Maxime Audinot, Fran{\c{c}}ois Bonnet, and Simon Viennot.
\newblock Optimal strategies against a random opponent in battleship.
\newblock In {\em The 19th Game Programming Workshop}, 2014.

\bibitem[CdFG20]{crombez2020efficient}
Lo{\"\i}c Crombez, Guilherme~D da~Fonseca, and Yan Gerard.
\newblock Efficient algorithms for battleship.
\newblock {\em arXiv preprint arXiv:2004.07354}, 2020.

\bibitem[Cle13]{clementis2013supervised}
Ladislav Clementis.
\newblock Supervised and reinforcement learning in neural network based
  approach to the battleship game strategy.
\newblock In {\em Nostradamus 2013: Prediction, Modeling and Analysis of
  Complex Systems}, pages 191--200. Springer, 2013.

\bibitem[GNR{\etalchar{+}}23]{geerling2023using}
Wayne Geerling, Kristopher Nagy, Elaine Rhee, Nicola Thomas, and Jadrian
  Wooten.
\newblock Using squid game to teach game theory.
\newblock {\em Journal of Economics Teaching}, 8(1):47--63, 2023.

\bibitem[Kan20]{kancko2020reinforcement}
Tom{\'a}{\v{s}} Kancko.
\newblock Reinforcement learning for the game of battleship.
\newblock {\em Masaryk University, Faculty of Informatics}, 2020.

\bibitem[RCH{\etalchar{+}}88]{rodin1988developing}
EY~Rodin, J~Cowley, K~Huck, S~Payne, and D~Politte.
\newblock Developing a strategy for ``battleship".
\newblock {\em MATH. COMP. MODEL.}, 10(2):145--153, 1988.

\end{thebibliography}

\end{document}